\documentclass[showpacs,preprintnumbers, amsmath,amssymb,APSl,prd,nofootinbib,superscriptaddress]{revtex4-2}
\usepackage{graphicx,color}
\usepackage{subfig}
\usepackage{amssymb}
\usepackage{hyperref}
\usepackage[utf8]{inputenc}
\usepackage[english]{babel}
\usepackage{epsfig}
\usepackage{wasysym}
\usepackage{color,xcolor}
\usepackage{amsmath}
\usepackage{bm}
\usepackage{epsfig}
\usepackage{amsfonts}
\usepackage{dcolumn}
\usepackage{float}

\hypersetup{colorlinks=true, linkcolor=blue, citecolor=green}

\begin{document}

\title{Gravitational lensing of a Schwarzschild-like black hole in Kalb-Ramond gravity}

	\author{Ednaldo L. B. Junior} \email{ednaldobarrosjr@gmail.com}
\affiliation{Faculdade de F\'{i}sica, Universidade Federal do Pará, Campus Universitário de Tucuruí, CEP: 68464-000, Tucuruí, Pará, Brazil}

     \author{José Tarciso S. S. Junior}
    \email{tarcisojunior17@gmail.com}
\affiliation{Faculdade de F\'{i}sica, Programa de P\'{o}s-Gradua\c{c}\~{a}o em F\'{i}sica, Universidade Federal do Par\'{a}, 66075-110, Bel\'{e}m, Par\'{a}, Brazill}

	\author{Francisco S. N. Lobo} \email{fslobo@ciencias.ulisboa.pt}
\affiliation{Instituto de Astrof\'{i}sica e Ci\^{e}ncias do Espa\c{c}o, Faculdade de Ci\^{e}ncias da Universidade de Lisboa, Edifício C8, Campo Grande, P-1749-016 Lisbon, Portugal}
\affiliation{Departamento de F\'{i}sica, Faculdade de Ci\^{e}ncias da Universidade de Lisboa, Edif\'{i}cio C8, Campo Grande, P-1749-016 Lisbon, Portugal}

    \author{\\Manuel E. Rodrigues} \email{esialg@gmail.com}
\affiliation{Faculdade de F\'{i}sica, Programa de P\'{o}s-Gradua\c{c}\~{a}o em F\'{i}sica, Universidade Federal do Par\'{a}, 66075-110, Bel\'{e}m, Par\'{a}, Brazill}
\affiliation{Faculdade de Ci\^{e}ncias Exatas e Tecnologia, Universidade Federal do Par\'{a}, Campus Universit\'{a}rio de Abaetetuba, 68440-000, Abaetetuba, Par\'{a}, Brazil}

    \author{Diego Rubiera-Garcia} \email{ drubiera@ucm.es}
\affiliation{Departamento de Física Téorica and IPARCOS, Universidad Complutense de Madrid, E-28040 Madrid, Spain}

\author{Luís F. Dias da Silva} 
\email{fc53497@alunos.fc.ul.pt}
\affiliation{Instituto de Astrof\'{i}sica e Ci\^{e}ncias do Espa\c{c}o, Faculdade de Ci\^{e}ncias da Universidade de Lisboa, Edifício C8, Campo Grande, P-1749-016 Lisbon, Portugal}

    \author{Henrique A. Vieira} \email{henriquefisica2017@gmail.com}
\affiliation{Faculdade de F\'{i}sica, Programa de P\'{o}s-Gradua\c{c}\~{a}o em F\'{i}sica, Universidade Federal do Par\'{a}, 66075-110, Bel\'{e}m, Par\'{a}, Brazill}

\begin{abstract}
In this paper, we investigate the gravitational lensing effect for the Schwarzschild-like black hole spacetime in the background of a Kalb-Ramond (KR) field proposed in [K.~Yang, {\it et. al.}, Phys. Rev. D \textbf{108} (2023) 124004]. The solution is characterized by a single extra parameter $l$, which is  associated to the Lorentz symmetry breaking induced by the KR field. First, we calculate the exact deflection angle of massive and massless particles for finite distances using elliptic integrals. Then we study this effect in the weak and strong field regimes, discussing the correction of the KR parameter on the coefficients of the expansions in both limits. We also find that increasing $l$ decreases the deflection angle. Furthermore, we use the available data from the Sagittarius $A^{\star}$ object, which is believed to be a supermassive black hole at the center of our galaxy, to calculate relevant observables, such as, the image position, luminosity, and delay time. The values found could be potentially measured in the weak field regime, though for strong fields one would have to wait for the next generation of interferometers.
\end{abstract}

\date{\today}

\maketitle

\section{Introduction}

More than a century after Einstein introduced the General Theory of Relativity (GR), we find ourselves at the dawn of a new epoch in gravitational physics, spurred by two landmark discoveries.
 In 2016, the LIGO and VIRGO Collaborations detected the first signals of gravitational waves, indicative of the merger of two black holes, which was later expanded to include the coalescence of a black hole and a neutron star \cite{LIGOScientific:2016aoc,LIGOScientific:2017vwq}. 
 On the other hand, in 2019, the Event Horizon Telescope (EHT) Collaboration unveiled the first-ever image of super-heated plasma swirling around the supermassive object residing at the core of the M87 galaxy \cite{fotoBN1,fotoBN2,fotoBN3,fotoBN4,fotoBN5,fotoBN6,fotoBN7,fotoBN8,fotoBN9}, later extending their observations to Sagittarius $A^{\star}$ (Srg $A^{\star}$) \cite{fotosrg1,fotosrg2,fotosrg3,fotosrg4,fotosrg5,fotosrg6}, the active radio-source at the nucleus of our Milky Way galaxy. Both observations affirm the presence of supermassive black holes.
Furthermore, the forthcoming advancements in Very Long Baseline Interferometry (VLBI), exemplified by projects like GRAVITY \cite{GRAVITY1,GRAVITY2,GRAVITY3,GRAVITY4,GRAVITY5,GRAVITY6}, are poised to scrutinize GR in its strong field regime with unprecedented precision. These developments will pave the way for exploring alternative gravitational theories beyond the tests presently carried out in the weak field limit \cite{Yunes:2013dva}.

These significant advancements stem from the paradigm shift introduced by GR in contrast to the Newtonian description of gravity, exemplified by its first observational test, namely, the deflection of light. Initially observed during the 1919 eclipse by the Eddington \cite{Dyson:1920cwa} and the Sobral \cite{Eclipse, Crispino:2020txj} expeditions, the bending of light rays by massive bodies manifests as gravitational lensing, a hallmark of how gravity intertwines the fabric of space-time with the motion of both massive and massless particles. Leveraging gravitational lensing, we have used stars, black holes, entire galaxies, and even galaxy clusters as natural telescopes, vastly expanding our capacity to observe remote celestial objects \cite{Congdon}. This achievement owes much to the pioneering work of Virbhadra and Ellis \cite{Virbhadra:1999nm}, who elucidated the gravitational lensing produced by a Schwarzschild black hole, introducing concepts such as the surface where the deflection angle tends to infinity and the phenomenon of multiple images. Generalizing their findings to encompass any spherically symmetric metric, this surface is termed the photon sphere \cite{Claudel:2000yi}, and plays a key role in generating images of compact celestial bodies. Indeed, Bozza \cite{Bozza:2002zj} proved that such a divergence is logarithmic and provided a method to compute observables of interest for any static and spherically symmetric metric. 

Various applications of this framework have been extensively explored in the scientific literature. It has been used in the context of supermassive black holes \cite{Bozza:2012by} and the trajectories of stellar bodies orbiting them \cite{Pietroni:2022cur}, as well as in elucidating the deflection of light by astrophysical compact objects \cite{Tsukamoto:2020iez,Tsukamoto:2020bjm,Tsukamoto:2021fsz,Zhang:2022nnj,Ghosh:2022mka,Nascimento:2020ime,Furtado:2020puz}, with overarching formalisms developed for such purposes \cite{Keeton:2005jd,Keeton:2006sa,Keeton:2006di}. Recent applications of gravitational lensing include observations of this phenomenon around supermassive black holes \cite{Ghaffarnejad:2014zva,Nightingale:2023ini,Legin:2022ovl}, microlensing's collective amplification of multiple images for detecting bodies that do not emit light, such as planets \cite{Wambsganss:1996he,Bozza:2018loy} or black holes \cite{Sajadian:2023xsf,Horvath:2011xr},  its use as a distance estimator \cite{Bozza:2003cp}, aiding in the study of binary star systems \cite{Shin:2012xz,Choi:2013ajr}, and its relevance in cosmological contexts \cite{Holz:1997ic,Lewis:2006fu,Frieman:2008sn,Poddar:2021sbc}. For a comprehensive overview of these methods and others employing weak lensing, we refer the reader to \cite{Weinberg:2013agg,AbhishekChowdhuri:2023ekr}. A major improvement of this formalism was introduced by  Ishihara {\it et al} in \cite{Ishihara:2016vdc} by extending the computation of the deflection angle to finite distances, as illustrated in \cite{DCarvalho:2021zpf,Carvalho:2021ajy} with the case of massive particles. 

The main aim of the present work is to investigate gravitational lensing for the  solution  presented in \cite{Yang:2023wtu}. This solution, a Schwarzschild-type metric, incorporates a spontaneous breaking of Lorentz symmetry through the non-minimal coupling of the gravitational field with the Kalb-Ramond field \cite{Kalb:1974yc}. 
While Lorentz symmetry stands as a cornerstone of physics, it is conjectured that under certain conditions, such as in the Standard Model \cite{Colladay:1996iz,Colladay:1998fq}, quantum gravity \cite{Alfaro:2001rb}, and other scenarios \cite{Jacobson:2000xp,Carroll:2001ws,Dubovsky:2004ud,Cohen:2006ky,Horava:2009uw,Bengochea:2008gz}, it may be violated. Previous studies have investigated gravitational lensing in theories featuring Lorentz-symmetry breaking, such as in \cite{DCarvalho:2021zpf,Ovgun:2018ran,Vagnozzi:2022moj}, and more recently, for alternative implementations of the Kalb-Ramond field \cite{Lessa:2019bgi,Kumar:2020hgm,Ditta:2023ccf}. 
In this work, we aim to derive exact expressions for gravitational lensing within this theory and apply them to both weak and strong field regimes. Additionally, we will extend the formalism of gravitational lensing to encompass finite distances to derive the corresponding expressions in such scenarios.

This work is organized as follows: In Sec. \ref{CG}, we introduce the Kalb-Ramond solution, review some of its properties, and determine the exact deflection angle of massive particles using elliptic integrals. In Sec. \ref{sec:weak}, we obtain an approximation for the deflection angle that is valid for the weak gravitational field regime, and in Sec. \ref{sec:strong} an approximation for the strong field regime. In Sec. \ref{sec:observaveis}, we use the coefficients obtained in the previous two sections to compute the observables associated with each limit. Finally, in Sec. \ref{sec:conclu} we draw our conclusions.

\section{Gravitational lensing in Kalb-Ramond theory}\label{CG}

\subsection{Kalb-Ramond field}

The Kalb-Rammond (KR) solution introduced in \cite{Yang:2023wtu} is a Schwarzschild-type solution that implements a breaking in Lorentz symmetry. It is given by the following line element
\begin{equation}
    ds^2 = -\left( \frac{1}{1-l} - \frac{2m}{r}\right) dt^2 +  \left( \frac{1}{1-l} - \frac{2m}{r}\right)^{-1}\,dr^2 + r^2 d \Omega^2\,,
    \label{eq:KRmetrica}
\end{equation}
where $l$ is a dimensionless parameter that characterizes the effect of Lorentz symmetry violation arising from the non-zero vacuum expectation value of the Kalb-Ramond (KR) field permeating spacetime. Solar System tests, such as the Shapiro time delay, light deflection, and Mercury's perihelion precession, constrain the parameter $l$ to the interval $-6.1 \times 10^{-13} < l < 2.8 \times 10^{-14}$ \cite{Yang:2023wtu}. However, on a much larger scale of mass as given by the observations of the Sgr A$^*$ radiosource, assumed to hide a supermassive black hole, at the center of our own Milky Way galaxy, the parameter $l$ would be constrained to the interval $-0.18502 < l <0.06093 $ \cite{Ednaldo}.
Table \ref{tab:tabela1} shows the estimated mass and distance from us of Sgr A$^*$ as measured by the Keck and VLTI.

\begin{table}[h]
\centering
\resizebox{0.5\textwidth}{!}{%
\def\arraystretch{1.4}
\begin{tabular}{cccc}
\hline \hline
\multicolumn{4}{c}{Parameter values}                                                                               \\ \hline
\hspace{1 mm}Survey\hspace{1 mm} & \hspace{1 mm}$M (\times 10^6 M_{\odot})$\hspace{1 mm} & \hspace{1 mm}$D$\hspace{1 mm} (kpc) & \hspace{1 mm}Reference\hspace{1 mm} \\ 
Keck                             & $3.951 \pm 0.047$                                     & $7.953 \pm 0.050 \pm 0.032$         & {\cite{Do:2019txf}} \\ \hline
\end{tabular}%
}
\caption{Sgr A* mass and distance as inferred by the Keck and VLTI.}
\label{tab:tabela1}
\end{table}

\subsection{Exact gravitational lensing in the KB theory \label{sec:lens}}

In this section, we obtain the exact analytical expression for the deflection angle in terms of elliptic integrals. To this end, let us start with the general static, spherically symmetric line element given by
\begin{equation} \label{eq:lineel}
    ds^2 = -A(r) dt^2 + B(r) dr^2 + C(r) d \Omega^2,
\end{equation}
where $d\Omega^2=d\theta^2 + \sin^2 \theta d\phi^2$ is the two-spheres line element. Though it is always possible to remove one of the coefficients $\{A(r),B(r),C(r)\}$ in favor of the other two and end up with just two independent functions, for the moment we shall work under this general form. The static and spherical symmetry allow to introduce two conserved quantities of motion, namely, $E=-A\dot{t}^2$ and $L=r^2 \dot{\phi}$, interpreted as the energy and angular momentum per unit mass, respectively. Here, an overdot denotes a derivative with respect to the affine parameter.

Under these conditions, the geodesic equation for a massive particle of velocity $v$ can be written as  \cite{DCarvalho:2021zpf} 
\begin{equation}
    \left( \frac{dr}{d \phi} \right)^2 = \frac{r^4}{A(r)B(r)} \left[  \frac{1}{b^2v^2} - \left( \frac{1-v^2}{b^2v^2} + \frac{1}{r^2} \right)A(r)^2  \right],
    \label{eq:motion}
\end{equation}
where $b \equiv L/E$ is the impact parameter. In the context of gravitational lensing, a particle will approach a certain closest distance $r_0$ to the source before being deflected by its gravitational influence. In terms of such a distance, the impact parameter can be written as
\begin{equation}
  b = \sqrt{\frac{C(r_0)}{A(r_0)}},
    \label{eq:b}
\end{equation}
which is the impact parameter of the trajectory defined in Eq. (\ref{eq:motion}) in terms of the distance of closest approach $r_0$. 

For the metric \eqref{eq:KRmetrica}, the geodesic equation  \eqref{eq:motion} can be suitably rewritten, using the change of variable $r=1/u$ as 
\begin{equation}
    \left( \frac{du}{d \phi} \right)^2 = 2m G(u),
\end{equation}
where the function  $G(u)$ is explicitly given by
\begin{equation} \label{eq:GuKB}
    G(u) =u^3+\frac{u^2}{2 (l-1) m}+\frac{u}{b^2}
   \left(\frac{1}{v^2}-1\right)+\frac{u^2}{2 (l-1) m}+\frac{l-v^2}{2 b^2 (l-1) m v^2}.
\end{equation}

For a source and an observer located at a finite distance, the deflection angle was computed by Ishihara \textit{et al} in \cite{Ishihara:2016vdc} as
\begin{equation}
    \alpha = \sqrt{\frac{1}{2M}} \left( \int_{u_R}^{u_0} \frac{du}{\sqrt{G(u)}}+\int_{u_S}^{u_0} \frac{du}{\sqrt{G(u)}} \right) + \Psi_R - \Psi_S.
    \label{eq:alphageral}
\end{equation}
In this expression $u_0 = 1/r_0$ is the inverse of the distance of closest approach, $u_R = 1/r_R$ and $u_S=1/r_S$ are the inverse of the radius of the observer (R) and the source (S), respectively, while the function $\Psi$ is given by
\begin{equation} \label{eq:phifun}
    \Psi(u) = \arcsin \left[ buv \sqrt{A(u)} \right],
\end{equation}
where $\Psi_R \equiv \Psi(u_R)$ and $\Psi_S \equiv \Psi(u_S)$  implement the finite distance corrections. In particular, when both observer and source are at infinity, $u_r=u_s \to 0$, then the well known expression of the deflection angle for infinite distances \cite{Bozza:2002zj} is recovered:
\begin{equation}
    \alpha_{\infty} = \sqrt{\frac{2}{M}}\int_{0}^{u_0} \frac{du}{\sqrt{G(u)}} - \pi,
\end{equation}

For the KR solution (\ref{eq:KRmetrica}) we compute the integrals appearing in the finite-distance deflection angle \eqref{eq:alphageral} as follows. We first assume the function $G(u)$ to be written as
\begin{equation}
    G(u) = (u-u_1)(u-u_2)(u-u_3),
\end{equation}
where $u_1,u_2$ and $u_3$ are the roots of  a cubic polynomial, and take $u_2=1/r_0$. Such roots can be found by comparing the expression with the actual one of the KR field, Eq. (\ref{eq:GuKB}), which after some algebra provides the relations
\begin{equation}
    \begin{aligned}
        & u_1 +u_3 +1/r_0 =   \frac{1}{2m(1-l)}, \\
        & u_1u_3 +u_1/r_0 +u_3/r_0 =  \frac{\frac{1}{v^2}-1}{b^2},
    \end{aligned}
\end{equation}
which can be solved to provide the roots $u_1$ and $u_3$ as
\begin{eqnarray}
    u_1 &=&-\frac{2 (l-1) m u_0+1}{4 (l-1)^2 m v^2}
     \left[
   \sqrt{-\frac{(l-1)^2 v^2 \left(16 (l-1) m^2 u_0^2
   \left(v^2-1\right)+6 (l-1) m u_0
   v^2-v^2\right)}{ (2 (l-1) m u_0+1)}}+(l-1)
   v^2\right],
   \label{eq:u1} \\
    u_3 &=& \frac{2 (l-1) m u_0+1}{4 (l-1)^2 m v^2} 
    \left[
   \sqrt{-\frac{(l-1)^2 v^2 \left(16 (l-1) m^2 u_0^2
   \left(v^2-1\right)+6 (l-1) m u_0
   v^2-v^2\right)}{ (2 (l-1) m u_0+1)}}-(l-1)
   v^2\right]\,,
   \label{eq:u3}
\end{eqnarray}
where the signs when taking the square roots have been chosen so as to have the roots ordered as  $u_3> u_2> u_1$. This way, both integrals appearing in Eq. \eqref{eq:alphageral} can be put under the form
\begin{equation}
    I_1(u^{\prime}) =  \int_{u^{\prime}}^{u_2}  \frac{du}{\sqrt{(u-u_1)(u-u_2)(u-u_3)}} \,,
\end{equation}
with $u_3>u_2>u^{\prime}>u_3$ and $u^{\prime}$ being any finite distance. This integral can be carried out with the help of elliptic functions as 
\begin{equation}
    I(u^{\prime}) = \frac{2}{\sqrt{u_3-u_1}} F(\delta,k^2),
\end{equation}
where $F(\delta,k^2) $ is a incomplete first-order elliptic integral and the constants take the expressions
\begin{equation}
\delta = \arcsin{\sqrt{\frac{(u_3-u_1)(u_2-u^{\prime})}{(u_2-u_1)(u_3-u^{\prime})}}} \,,      \qquad
k^2 = \frac{u_2-u_1}{u_3-u_1},
\end{equation}

On the other hand, to find the finite-distance correction integrals $\Psi_R$ and $\Psi_S$ we just need to write Eq.(\ref{eq:phifun}) explicitly for the KR metric (\ref{eq:KRmetrica}) as 
\begin{equation}
    \Psi(u) = \sin ^{-1}\left(b u v \sqrt{\frac{\frac{1}{1-l}-2 m u}{1-\frac{1}{1-l}+\frac{2 m}{1-v^2}}}\right).
    \label{eq:Psi}
\end{equation}
evaluate at the desired finite distances $u_R$ and $u_S$. 

Collecting all the above expressions, we find the deflection angle in the finite-distance regime as 
\begin{equation}
    \alpha (r_0) = \sqrt{\frac{2}{(u_3-u_1)M}} \left(  F(\delta(u_R),k)+F(\delta(u_S),k) \right) +  \Psi_R - \Psi_S
    \label{eq:alphar0}
\end{equation}
Obviously, in the infinite-distance limit, in which $u_R =u_S \to 0$ the pieces in $\Psi_R$ and $\Psi_S$ vanishes, and we collect the infinite-distance deflection angle (and the one of Schwarzschild case if we set $l \to 0$). Alternatively, the deflection angle can be rewritten in terms of the impact parameter $b$. To this end, we note that Eq. (\ref{eq:b}) for the KR field (\ref{eq:KRmetrica})  becomes 
\begin{equation}
    \frac{b}{r_0} = \sqrt{\frac{r_0-l r_0}{2 (l-1) m+r_0}},
    \label{eq:bero}
\end{equation}
which is an implicit equation for $r_0$ and can be solved as 
\begin{equation}
    \frac{r_0}{b} = 2  \sqrt{\frac{1}{3-3 l}} \cos \left[\frac{1}{3} \cos
   ^{-1}\left(-\frac{3 \sqrt{3} (1-l)^{3/2} m}{b}\right)\right].
   \label{eq:beer0}
\end{equation}
This facilitates the expression of $\alpha(r_0(b))$ by substituting this equation into the aforementioned expressions. In the subsequent analysis, we will divide our analysis into weak and strong field regimes.

\section{Weak field approximation \label{sec:weak}}

In this section, we compute the deflection angle  in the weak field limit. In this regime, we assume that both the source and the observer are very far from the lens and the light rays are only slightly distorted by the lens. To start with, we define the lens equation as \cite{Virbhadra:1999nm}
\begin{equation}
    \tan \beta = \tan \theta- D \left( \tan \theta + \tan(\alpha - \theta)  \right ),
    \label{eq:lensequation}
\end{equation}
where \textbf{$\beta$} and $\theta$ are the angular position of the source and the lensed images, respectively, and
\begin{equation}
    D = \frac{D_{LS}}{D_{OS}} \,.
\end{equation}
$D_{OS}= D_{LS} +D_{OL}$ is the distance between the observer and the source, with $D_{LS}$ and $D_{OL}$ the distance between the lens and the source and the lens and the observer. 

For a static, spherically symmetric metric of the general form (\ref{eq:lineel}), the deflection angle, assuming both observer and source to be located at infinity, can be computed using the standard formula \cite{Virbhadra:1998dy,Congdon}
\begin{equation}
    \alpha (r_0) = 2 \int_{r_0}^{\infty} \left| \frac{d \phi}{dr}\right| - \pi.
    \label{eq:alphain}
\end{equation}

We follow the approach introduced in \cite{Keeton:2005jd} by which the result of the above integral can be approximated by a series of the following form
\begin{equation}
    \alpha (b) = A_1 \left(  \frac{m}{b}\right) + A_2 \left(  \frac{m}{b}\right)^2 + \mathcal{O} \left(  \frac{m}{b}\right)^3.
    \label{eq:alphau}
\end{equation}
Here, the deflection angle is written as a function of the impact parameter $b$, since it is a gauge invariant variable (while the closest approach distance has a gauge dependence). The $A_i$ are coefficients to be calculated, which can be simple numbers or depend on a parameter of the solution. Using Eq. (\ref{eq:motion}) for the KR metric (\ref{eq:KRmetrica}), we are driven to solve the integral
\begin{equation}
    \alpha(r_0) = \int_{0}^{1} -\frac{2dx}{\sqrt{\frac{\frac{1}{1-l}-2 h}{v}-\left(\frac{1}{1-l}-2
   h x\right) \left(\left(\frac{1}{v^2}-1\right)
   \left(\frac{1}{1-l}-2 h\right)+x^2\right)}},
   \label{eq:inth}
\end{equation}
where $h=m/r_0$ and $x = r_0/r$. We then use a Taylor series expansion for the variable $h$ so the result becomes an expansion in $m/r_0$ which subsequently is written in terms of $m/b$ using yet another expansion in Eq. \eqref{eq:beer0}
\begin{equation}
    r_0 = b \left[\sqrt{\frac{1}{1-l}} -\frac{ (1-l)
   m}{b} -\frac{3
    (1-l)^{5/2} m^2}{2
   b^2} -\frac{4  (1-l)^{4} m^3}{b^3} \right].
\end{equation}
Thus, the deflection angle reads as in Eq.(\ref{eq:alphau}) with the coefficients 
\begin{equation}
    \begin{aligned}
        A_1 = \frac{-2  (1-l)^{5/2}}{(l v+1) (v (l-v-1)+1)} \Biggl[ c_1 +c_1(3l-2)v  
         +v^2 \left( \frac{l \left(c_2-2 c_3\right)+c_3}{(l-1) v} \right)  \\
        + v^3 \left( \frac{l \left(2 l \left(c_3-c_2\right)+2 c_2-3 c_3\right)+c_2}{(l-1) v} \right)    \Biggr],
    \end{aligned}
    \label{eq:A1}
\end{equation}
and 
\begin{eqnarray}
         A_2 &=&  \frac{3 (l-1)^{5/2} (v (5 l-v-5)+1) \left((l-1)^2 v^2 \left(2\log c_2+\log (l-1)\right)-2 \log \left((l-1) v+\sqrt{l-1} c_3\right)\right)}{4 v^2} 
         \nonumber \\
        && +\frac{(l-1)^4}{2 (l v+1)^2(c_2)^2} \Biggl[  -\frac{3 c_3}{v-l v} +v \left( \frac{(21 l-1) c_3}{(l-1) v}\right)
        -(1-l) v^2 \left(  \frac{11 (3 l-1) c_3}{(l-1) v}   \right) + v^4 \left( \frac{(5 (2-3 l) l+2) c_3}{(l-1) v}\right) 
        \nonumber \\
       && +v^3 \left(       \frac{15 l^4 v c_3-42 l^3 v c_3+l^2 v \left(8 c_2+25
   c_3\right)+2 l v \left(9 c_3-4 c_2\right)-8 l
   \left(c_2+c_3\right)-16 v c_3}{(l-1)^2 v^2} \right) \Biggr]\,,
    \label{eq:A2}
\end{eqnarray}
and where we have introduced the constants
\begin{equation}
   c_1 = \frac{v}{c_2+c_3} \,, \qquad 
   c_2 = \sqrt{ v^2 +v(1-l)-1} \,, \qquad
   c_3 = \sqrt{(v-1) (l v+1)}.
\end{equation}
These expressions recover the known ones of the Schwarzschild case when $l=0$ and $v=1$ \cite{Keeton:2005jd}. Now, let us refine the preceding result for the finite-distance scenario.

For a finite-distance case, using the expressions found so far and Taylor series expansion allows to integrate the deflection angle integral (\ref{eq:alphageral}) as
\begin{eqnarray}
     \alpha &=& \frac{(l-1) m \left(v^2 \left(b^2 u_R^2-1\right)+2 l-1\right)-b v \sqrt{v^2 \left(b^2
   u_R^2-1\right)+l} \sinh ^{-1}\left(\frac{b u_R v}{\sqrt{l-v^2}}\right)}{b v \sqrt{\frac{v^2
   \left(b^2 u_R^2-1\right)+l}{l-1}}}
   \nonumber \\
   &&
   +\frac{b (l-1) l \left(-\frac{1}{l}\right)^{5/2} m u_R v^2
   \left(l u_R \left(v^2-1\right)+1\right)}{\left(v^2-1\right) \sqrt{\frac{b^2 u_R^2
   v^4}{l}+1}}
   +\frac{b (l-1) \left(-\frac{1}{l}\right)^{3/2} m u_S v^2
   \left(l u_S \left(v^2-1\right)+1\right)}{\left(v^2-1\right) \sqrt{\frac{b^2 u_S^2
   v^4}{l}+1}}
    \nonumber \\
   &&
   +\frac{(l-1) m \left(v^2 \left(b^2 u_S^2-1\right)+2 l-1\right)-b v \sqrt{v^2 \left(b^2
   u_S^2-1\right)+l} \sinh ^{-1}\left(\frac{b u_S v}{\sqrt{l-v^2}}\right)}{b v \sqrt{\frac{v^2
   \left(b^2 u_S^2-1\right)+l}{l-1}}}
      \nonumber \\
    &&
   +\sin ^{-1}\left(b \sqrt{-\frac{1}{l}} u_R v^2\right)-\sin ^{-1}\left(b
   \sqrt{-\frac{1}{l}} u_S v^2\right) + \mathcal{O} \left( \frac{m}{b} \right)^2 \,,
   \label{eq:alphawf2}
\end{eqnarray}
which provides the value $4 m/b$ of the Schwarzschild geometry if $l=0$, $v=1$ and $u_S = u_R = 0$.

\section{Strong deflection angle \label{sec:strong}}

\subsection{General approach}

In the previous section, we used a formalism that applies to the case where the closest approach distance is much larger as compared to the mass of the lens (in our case, the black hole). We now turn our attention to the computation of the opposite end of gravitational lensing, namely, the strong deflection regime. In such a regime one considers the trajectories of massless particles that get very close to the last unstable orbit. This is the closest orbit any photon can get to the black hole before getting captured by it to eventually be swallowed up by its event horizon. The corresponding radius $x_m$ is given by the resolution of the following equation 
\cite{Claudel:2000yi,Virbhadra:2002ju}
\begin{equation}
    \frac{C^{\prime}(x)}{C(x)} = \frac{A^{\prime}(x)}{A(x)}\,,
    \label{eq:fotonesfera}
\end{equation}
and which defines the so-called {\it photon sphere} of the black hole. A photon which backtracked from the observer's screen towards the photon sphere would formally have an infinite deflection angle. The corresponding value of the impact parameter $b=b_m$ for the radius $x_m$ is dubbed as the critical impact parameter.

In order to compute the deflection angle in this strong field regime, we follow the standard formalism developed by  Bozza \cite{Bozza:2002zj}. In the limit $b \to b_m$, the deflection angle $\alpha$ can be approximated by a logarithmic expansion of the form
\begin{equation}
    \alpha (b) = b_1 \log \left( \frac{b}{b_m} - 1 \right) + b_2 + \mathcal{O} \left( b - b_m \right).
    \label{eq:explog}
\end{equation}
Here $b_1$, $b_2$, together with $b_m$, are coefficients uniquely depending on the background geometry, and thus this approach nicely connects the geometrical properties with the motion of light rays.

We will briefly discuss how to obtain the other two coefficients. We emphasize that the starting point is the same as before, namely to propose an approximate result for the integral \eqref{eq:alphageral}. First, we define the variables
\begin{equation}
    y = A(x) \, , \qquad
    \zeta = \frac{y - y_0}{1 - y_0},
\end{equation}
where $y_0 = A(x_0)$.  This leads to the expression of the deflection angle via the integral
\begin{equation}
    \alpha (x_0) = I(x_0) - \pi \,, \qquad
    I(x_0) = \int_0^1 R( \zeta, x_0) f(\zeta, x_0) d \zeta \,,
    \label{eq:Itotal}
\end{equation}
where the function $R( \zeta, x_0)$ is given by
\begin{equation}
    R( \zeta, x_0) = \frac{2 \sqrt{By}}{C A^{\prime}} \left( 1 - y_0 \right) C_0, 
\end{equation}
and it is regular for every value of  $\zeta$ and $x_0$. The function $f(\zeta, x_0)$ reads
\begin{equation}
    f(\zeta, x_0) = \frac{1}{\sqrt{y_0 - \bigl[ (1-y_0) \zeta + y_0\bigr]\frac{C_0}{C}}},
    \label{eq:f01}
\end{equation}
and has a divergence for $\zeta \rightarrow 0$.  All functions without the subscript 0 are evaluated
at $x = A^{-1}\bigl[(1 - y_0) \zeta + y_0 \bigr]$. 

We now approximate the function $f(\zeta, x_0)$  in this strong field regime as
\begin{equation}
    f(\zeta, x_0) \sim f_0(\zeta, x_0) = \frac{1}{\sqrt{\beta_1 \zeta + \beta_2 \zeta^2}},
    \label{eq:f02}
\end{equation}
with the constants
\begin{eqnarray}
    \beta_1 &=& \frac{1 - y_0}{C_0 A^{\prime}_0} \left(  C^{\prime}_0 y_0 - C_0 A^{\prime}_0  \right),
    \label{eq:beta1} \\
    \beta_2 &=& \frac{(1-y_0)^2}{2C_0^2 A^{\prime \ 3}_0}  \bigr[  2 C_0 C^{\prime}_0 A^{\prime \ 2}_0 +(C_0 C_0^{\prime \prime} - 2C_0^{\prime \ 2} )y_0 A_0^{\prime} - C_0 C_0^{\prime}y_0A_0^{\prime \prime}  \bigl].
    \label{eq:beta2}
\end{eqnarray}
From Eq. \eqref{eq:f02} we can see that if $\beta_1 \neq 0$, the leading order of divergence in \eqref{eq:f01} is $\zeta^{-1/2}$, while if $\beta_1 = 0$ the divergence is $\zeta^{-1}$. In the first case $f_0$ can be integrated and the result is finite, while in the second case the integral diverges. Returning to the original variables, we note that $\beta_1$ vanishes at the photon sphere location, $x_0 = x_m$. 

The canonical approach to deal with this problem is to split the integral \eqref{eq:Itotal} in two pieces as
\begin{equation}
    I(x_0) = I_D (x_0) + I_R (x_0),
\end{equation}
where
\begin{equation}
    I_D (x_0) = \int_0^1 R( 0, x_m) f_0(\zeta, x_0) d \zeta,
\end{equation}
is the divergent part and
\begin{equation}
    I_R (x_0) = \int_{0}^1 g(\zeta, x_0) d \zeta,
\end{equation}
is the regular one, with
\begin{equation}
    g(\zeta, x_0) = R( \zeta, x_0) f(\zeta, x_0) - R( 0, x_m) f_0(\zeta, x_0).
\end{equation}
This way, one can compute the logarithmic approximation to the deflection angle as \cite{Bozza:2002zj}
\begin{equation}
    \alpha (x_0) =  - \left( \frac{R(0,x_m)}{\sqrt{\beta_{2m}}}\right) \log \left( \frac{x_0}{x_m} -1 \right) +  \frac{R(0,x_m)}{\sqrt{\beta_{2m}}} \log \frac{2(1- y_m)}{A^{\prime}_m x_m} +\int_0^1 g(\zeta, x_m) d \zeta- \pi + \mathcal{O} \left( x_0 - x_m\right).
   \label{eq:alphax0strong}
\end{equation}
where functions with index $m$ are calculated in  $x_0 = x_m$.  

As stated in the previous section, it is convenient to write this result in terms of the gauge invariant coordinate $b$. We can expand Eq. \eqref{eq:b} and write
\begin{equation}
    b - b_m = \beta_{2m} \sqrt{\frac{y_m}{C_m^3}} \frac{C^{\prime \ 2}_m }{2(1-y_m^2)} \left( x_0 - x_m \right)^2.
    \label{uex0}
\end{equation}
With the above equation we can write Eq. \eqref{eq:alphax0strong} in the form of Eq. \eqref{eq:explog}, where the coefficients $b_1$ and $b_2$ are
\begin{eqnarray}
    b_1 &=& \frac{R(0,x_m)}{2\sqrt{\beta_{2m}}},
    \label{eq:b1g} \\
    b_2 &=& \int_0^1 g(\zeta, x_m) d \zeta +  b_1 \log \frac{2 \beta_{2m}}{y_m} - \pi.
    \label{eq:b2g}
\end{eqnarray}

\subsection{KR geometries}

Let us now particularize the expressions above for the KR geometry (\ref{eq:KRmetrica}), namely, $A(x) = \frac{1}{1-l}- \frac{1}{x}$,  $B(x) = A(x)^{-1}$ and $C(x) = x^2$. For this geometry, the photon sphere condition (\ref{eq:fotonesfera}) reads
\begin{equation}
    x_m = \frac{3}{2}(1-l),
\end{equation}
and by substituting this result into Eq. \eqref{eq:b} we get 
\begin{equation}
  b_m =   \frac{3}{2} \sqrt{3} (1-l)^{3/2} \,.
  \label{eq:bmsf}
\end{equation}

The functions $R(\zeta, x_m)$, $f(\zeta, x_m)$, and $g(\zeta,x_m)$ in this case become
\begin{equation}
    \begin{aligned}
       & R(\zeta,x_m) = 3 \left(1-\frac{1}{3-3 l}\right) (l-1), \\
       &f(\zeta, x_m) = \frac{2 \sqrt{3}}{(2-3 l)\sqrt{\frac{ \zeta^3(3l-2)+3\zeta^2}{1-l}}},\\
         &g(\zeta,x_m) = -\frac{2 \left(\sqrt{\frac{-3 l \zeta+2 \zeta-3}{l-1}}+\sqrt{3}
   \sqrt{\frac{1}{1-l}}\right)}{\sqrt{\frac{1}{1-l}} \zeta
   \sqrt{\frac{(2-3 l) \zeta-3}{l-1}}},
       \label{eq:fersc}
    \end{aligned}
\end{equation}
With the above expressions, one finds the strong deflection coefficients in this KR case as
\begin{eqnarray}
    b_1 &=& \frac{1}{\sqrt{\frac{1}{1-l}}}.
    \label{eq:b1sf} \\
    b_2 &=& \frac{-\pi  \sqrt{-9 l+\frac{1}{1-l}+3}+2 \log \left((2-3
   l)^2\right)-3 l \log \left(\frac{3}{2} (2-3
   l)^2\right)+\log \left(\frac{9}{4}\right)}{\sqrt{-9
   l+\frac{1}{1-l}+3}}
	 \nonumber \\
   &&+\frac{-2 \log (2-3 l)-4 \tanh
   ^{-1}\left(\sqrt{l+\frac{1}{3}}\right)+\log
   (144)}{\sqrt{\frac{1}{1-l}}}
   \label{eq:b2sf}
\end{eqnarray} 

\begin{figure}[t!]
    \centering
   \includegraphics[scale=0.8]{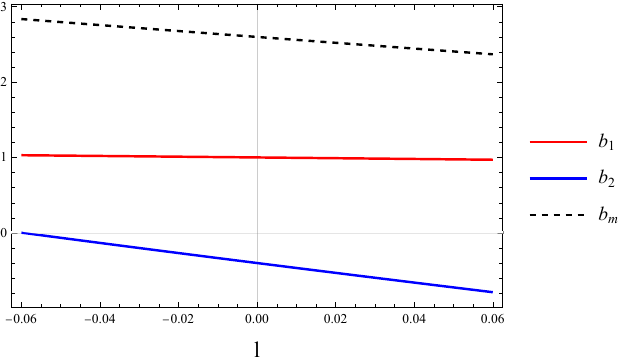}
   \includegraphics[scale=0.8]{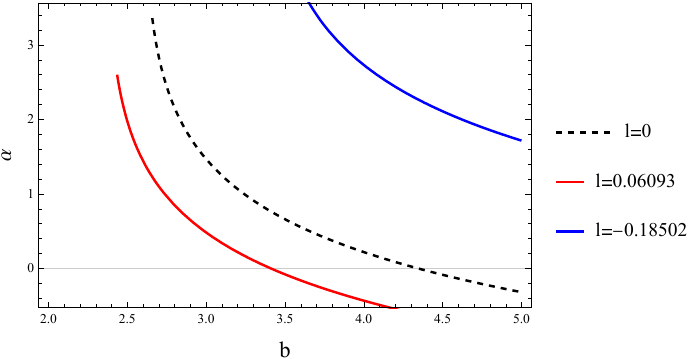}
    \caption{Let plot: the coefficients of the logarithmic expansion $b_1, b_2$ and the critical impact parameter $b_m$  as a function of the parameter $l$. Right plot: the deflection angle $\alpha$ as a function of the impact parameter $b$ for three values of the parameter $l$. }
    \label{fig:coef}
\end{figure}

In Fig. \ref{fig:coef} (left) we depict the strong field deflection coefficients $b_1$ and $b_2$, as well as the critical impact parameter $b_m$, as a function of the KR parameter $l$. We see that $b_2$ and $b_m$ decrease with increasing $l$, while $b_1$ remains practically constant. These coefficients allow to compute the strong deflection angle, which is depicted in Fig.  \ref{fig:coef} (right)  as a function of the impact parameter, for two values of $l$ and compared to the Schwarzschild geometry. We see that positive (negative) values of $l$ bend space-time more (less) than in the Schwarszschild case, a reflection of the variation of the parameters $b_2$ and $b_m$ with the KR field $l$.

\section{Observables} \label{sec:observaveis}

In this section, we calculate the observables associated with gravitational lensing.  Although only the positions of the lens, the source, and the image, as well as the brightness of the image, are apparent, we know that a single source may yield multiple images in the strong field regime, yet not all of them will be observable. Likewise, the weak field limit has its associated observables. Therefore, we will split our analysis into each regime.

\subsection{Observables in the weak field regime \label{sec:obsevavelwf}}

We first compute the observables of the weak-field regime using the coefficients $A_1$ and $A_2$ derived in Eqs. (\ref{eq:A1}) and (\ref{eq:A2}). Such observables are the angular separation $P_t$, the difference of angular positions $\Delta P$, the total flux $F_t$, the difference of fluxes $\Delta F$, the centroid $\Theta_{cent}$ and the differential time delay $\Delta \tau$ of the images. These coefficients are defined in \cite{Keeton:2005jd} as
\begin{eqnarray}
    P_t &=& \theta^{+}+\theta^{-} = \sqrt{A_1
   \theta_E^2+\beta^2} + \frac{A_2 \theta_E \epsilon }{A_1} +\mathcal{O}(\epsilon)^2,
   \label{eq:P_t}
   \\
    \Delta P &=& \theta^{+}-\theta^{-} = |\beta| - \frac{A_2\theta_E |\beta|}{A_1 \sqrt{A_1
   \theta_E^2+\beta^2}} \epsilon +  \mathcal{O}(\epsilon)^2,
   \label{eq:deltap}
   \\
    F_t &=& F^{+} + F^{- } = \frac{F_{src} \left(A_1 \theta _E{}^2+2 \beta ^2\right)}{2 |\beta|  \sqrt{A_1 \theta _E{}^2+\beta ^2}},
    \label{eq:fluxt}
\\
     \Delta F &=& F^{+}- F^{-} =  F_{src}-\frac{A_2 F_{src} \theta _E{}^3}{2 \left(A_1 \theta _E{}^2+\beta ^2\right){}^{3/2}}\epsilon
     \label{eq:deltaf}
     \\
    \Theta_{cent} &=& \frac{\vartheta^{+}F^{+} - \vartheta^{-}F^{-} }{F_t} = \frac{|\beta|  \left(3 A_1 \theta _E{}^2+4 \beta ^2\right)}{2 A_1 \theta _E{}^2+4 \beta ^2}
    \label{eq:thetacent}
    \\
    \Delta \tau &=& \frac{D_{OL} D_{OS}}{c D_{LS}} \left(\frac{1}{2} |\beta|  \sqrt{A_1 \theta _E{}^2+\beta ^2}+\frac{1}{4} A_1 \theta _E{}^2 \ln
   \left(\frac{\sqrt{A_1 \theta _E{}^2+\beta ^2}+|\beta| }{\sqrt{A_1 \theta _E{}^2+\beta ^2}-|\beta| }\right)+\frac{|\beta |
     A_2 \theta _E}{A_1} \epsilon\right)\,,
     \label{eq:deltatau}
\end{eqnarray}
where the angular Einstein radius is defined as
\begin{equation}
    \theta_{E} = \sqrt{\frac{4 G m D_{LS}}{c^2D_{OL}D_{OS}}} = \sqrt{\frac{4  m D_{LS}}{D_{OL}D_{OS}}},
\end{equation}
and the dimensionless perturbation parameter is given by
\begin{equation}
    \epsilon =\frac{\tan^{-1}(m/D_{OL})}{\theta_{E}} = \frac{\theta_{E}}{4 D}.
    \label{eq:novosangulos}
\end{equation}

The radio-source at the center of our galaxy, Sagittarius $A^{\star}$ is believed to be a supermassive black hole. Table \ref{tab:tabela1} reports the estimated data on its mass and distance. 
Considering only nominal values, we have $\theta_E = 0.022 (D_{LS})^{1/2}$ arc s and $\epsilon = 1.9\times 10^{-4} \times (D_{LS})^{-1/2}$. The distance from the source to the $D_{LS}$ lens is typically of the order of 1 parsec, so we shall adopt this value. Taking these values we can conclude that the observables lie within the limits of the precision of today's measuring instruments \cite{Keeton:2006sa}. However, we do not know whether the difference in the observables between the Schwarzschild and the Kalb-Ramond solutions is also within this measurement capacity. To investigate this issue, in the figures given below, we assume three curves with the values for the $l$ parameter (determined in \cite{Ednaldo}), corresponding to $l=0$ (dashed), $l=-0.18502$ (green) and $l=0.06093$ (red) in the plots. In addition, the observed values also depend on the dimensionless parameter $\beta$. In order to generate our results we take values of $0<\beta<0.1$, a range in which we already find significant variations. 

\begin{figure}[t!]
    \centering
   \includegraphics[scale=0.9]{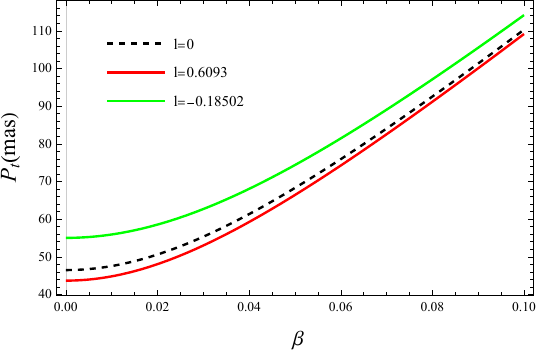}
   \includegraphics[scale=0.9]{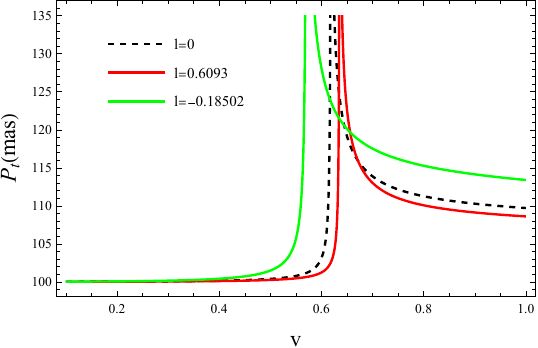}
    \caption{Left plot: $P_t$ as a function of the variable $\beta$ for three values of $l$. Right plot: $P_t$ as a function of the variable $v$ for three values of $l$.}
    \label{fig:thetamais1}
\end{figure}

In Fig. \ref{fig:thetamais1} we depict  $P_t$ as a function of $\beta$ for a fixed velocity $v=0.9$, finding a small difference between the curves and that as $l$ increases, the angular separation decreases; $P_t$ as a function of velocity $v$ for $\beta =0.1$, finding that the difference between the curves only occurs from $v \sim 0.6$.

\begin{figure}[t!]
    \centering
   \includegraphics[scale=0.9]{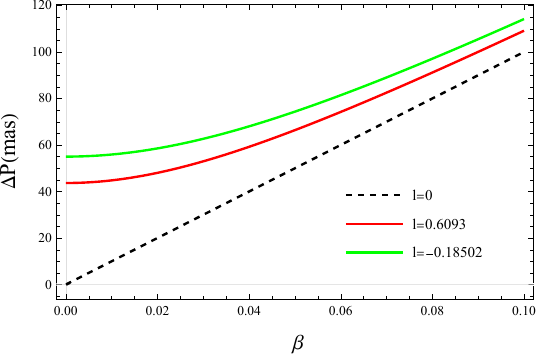}
   \includegraphics[scale=0.9]{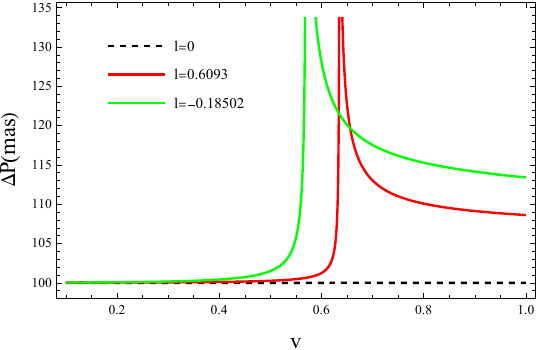}
    \caption{Left plot:  $\Delta P$ as a function of the variable $\beta$. Right plot: $\Delta P$ as a function of the variable $v$ for three values of $l$.}
    \label{fig:thetamenos}
\end{figure}

Figure \ref{fig:thetamenos}  shows $\Delta P$ as a function of $\beta$ and $v$, respectively. We see that the angular difference is smaller compared to the two extreme $l$ values and does not change with the variation of velocity. 

\begin{figure}[t!]
    \centering
   \includegraphics[scale=0.9]{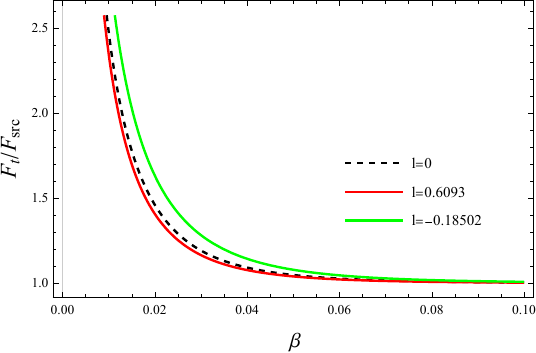}
   \includegraphics[scale=0.9]{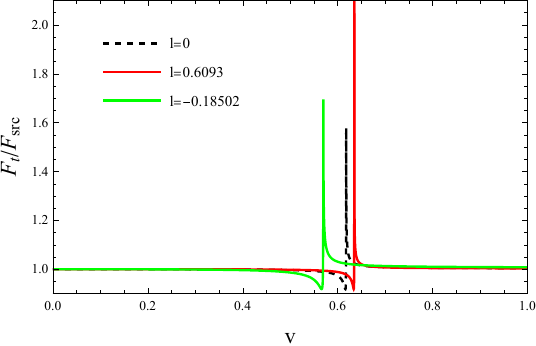}
    \caption{Left plot: $F_t$ as a function of the variable $\beta$ for three values of $l$. Right plot: $F_t$ as a function of the variable $v$ for three values of $l$.}
    \label{fig:Ft1}
\end{figure}

The total flux $F_t$ is shown in Fig. \ref{fig:Ft1}. This quantity is related to the magnification as each flux follows $F_i = |\mu_i| F_{src}$, where $F_{src}$ is the flux from the source. We see that for $\beta$ close to zero, the total flux tends to infinity and approaches the flux from the source as $\beta$ increases. We also see that the variation of velocity $F_t$ does not change, except for $v \approx 2/3$, and the  values of $l$ differ only slightly. 

\begin{figure}[t!]
    \centering
   \includegraphics[scale=0.9]{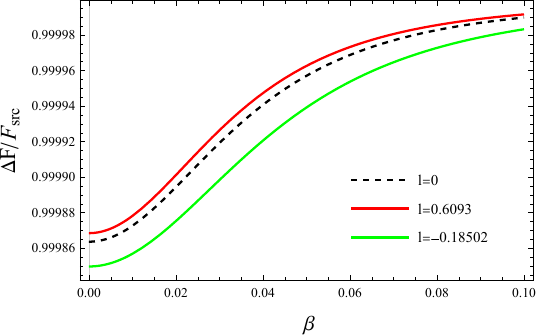}
   \includegraphics[scale=0.9]{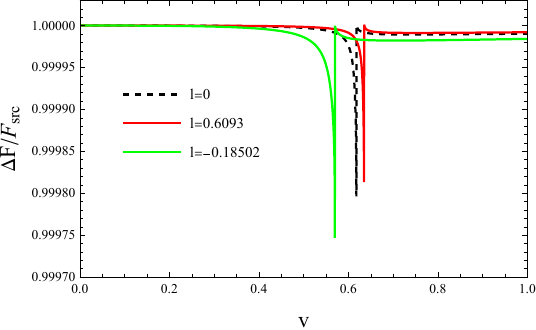}
    \caption{Left plot: $\Delta F$ as a function of the variable $\beta$ for three values of $l$. Right plot: $\Delta F$ as a function of the variable $v$ for three values of $l$.}
    \label{fig:deltaf}
\end{figure}

The difference in flux $\Delta F$ is shown in Fig. \ref{fig:deltaf}, and has similar properties to the total flux. The only change is that $\Delta F$ is higher for $l=0.6093$ and lower for $l=-0.18502$, i.e. the reverse occurs.

\begin{figure}[t!]
    \centering
   \includegraphics[scale=1]{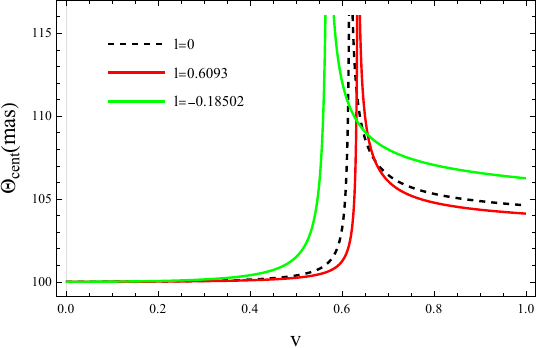}
    \caption{Graphical representation of the function $\Theta_{cent}$ as a function of the variable $v$ for three values of $l$.}
    \label{fig:tempo}
\end{figure}

In Fig. \ref{fig:tempo} we show the centroid $\Theta_{cent}$ as a function of the speed $v$. We see that this observable is higher as $l$ decreases, and the difference between the three $l$ curves becomes clearer for $v>0.5$. 

\begin{figure}[t!]
    \centering
   \includegraphics[scale=0.9]{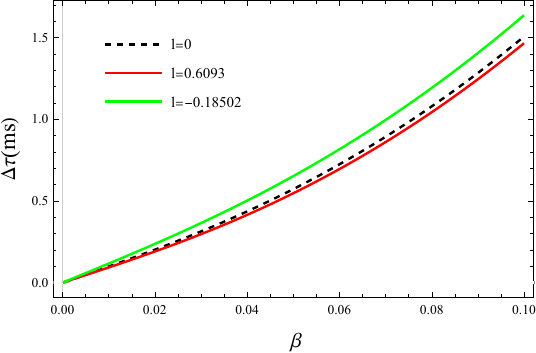}
   \includegraphics[scale=0.9]{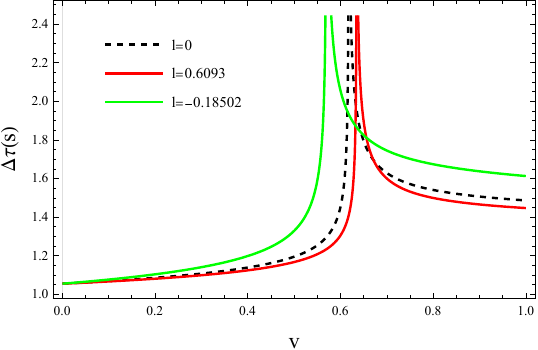}
    \caption{Left plot: $\Delta \tau$ as a function of the variable $\beta$ for three values of $l$. Right plot:  $\Delta \tau$ as a function of the variable $v$ for three values of $l$.}
    \label{fig:tempo2}
\end{figure}

Figure \ref{fig:tempo2} shows the time delay $\Delta \tau$. In the left plot we see that it gets higher with increasing $\beta$. However, the distance between the $l$ curves is in the order of milliseconds. In addition, $\Delta \tau$ is larger for the smallest value of $l$. In the right plot we see that the time delay increases until $v \approx 2/3$ and then decreases. 

As we have already said, all the values we find here for these observables are theoretically measurable with today's instruments \cite{Keeton:2006sa}. However,  disentangling the differences between the Schwarzschild and the Kalb-Rammond solutions is probably beyond the their reach. Regarding this, we also point out the difficulty in finding a source that is aligned with the lens.

\subsection{Observables in the strong field regime}

\begin{figure}[t!]
    \centering
   \includegraphics[scale=0.9]{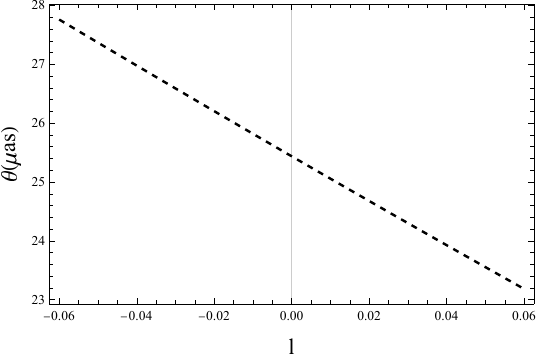}
   \includegraphics[scale=0.9]{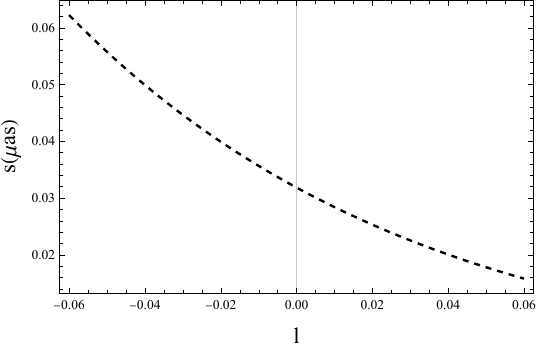}
   \includegraphics[scale=0.9]{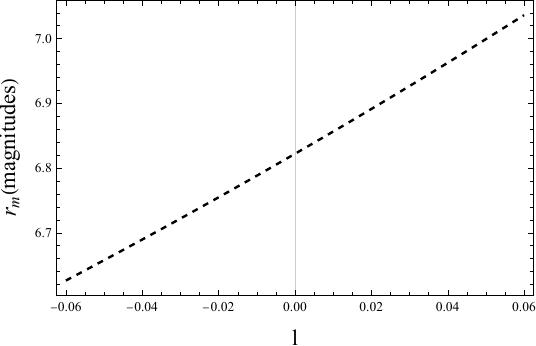}
   \includegraphics[scale=0.9]{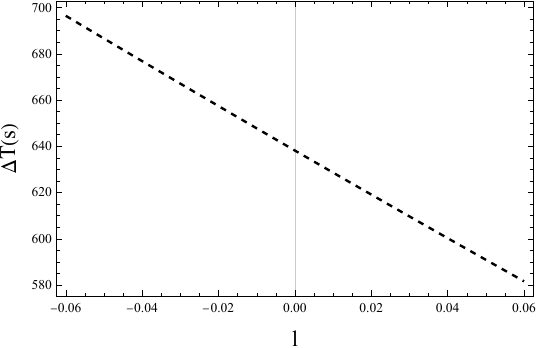}
    \caption{From  left to right and bottom to top: graphical representation of $\{\vartheta_{\infty},s,r_m,\Delta T_{2,1}\}$, as given by Eqs. (\ref{eq:thetainfity}), (\ref{eq:s}), (\ref{eq:r_m}), (\ref{eq:delaysf}), respectively, as functions of $l$.}
    \label{fig:thetainfi}
\end{figure}

Within the strong gravitational lensing regime, one expects multiples images to be formed when the impact parameter reaches the capture radius. However, higher-order images beyond the main one cannot be typically resolved, at least with today's technological capabilities. Instead of working with several individual images, V. Bozza \cite{Bozza:2002zj,Bozza:2003cp} describes the observation values in this limiting case, in which only the first image is fully resolved individually and the others are observed as a group. Using this fact, we focus on the following observables: asymptotic position approached by a set of images $\vartheta_{\infty}$, the distance between the first image (labeled $\vartheta_1$) and the others $s$,  the ratio between the flux of the first image and the flux of all the other images $r_m$, and in the time delay between one photon with two loops from one photon with one loop around the lens. These are given by
\begin{eqnarray}
    \vartheta_{\infty} &=& \frac{b_m}{ D_{OL}},
    \label{eq:thetainfity} \\
    s &=& \vartheta_1 - \vartheta_{\infty}  = \vartheta_{\infty} e^{\frac{b_2-2\pi}{b_1} },
    \label{eq:s}\\
    r_m &=& e^{\frac{2\pi}{b_1}},
    \label{eq:r_m}  \\
    \Delta T_{2,1} &=& [2 \pi -2\gamma] b_m + 2 \sqrt{\frac{B_m}{A_m}} \sqrt{\frac{b_m}{c_1}}e^{\frac{b_2}{2b_1}} \left(  e^{- \frac{ \pi \gamma}{b_1}}-  e^{- \frac{2 \pi \gamma}{b_1}}\right).
    \label{eq:delaysf}
\end{eqnarray}
which are functions of the strong-field coefficients and, in addition, for the time delay quantity also on the coefficient $\gamma$ which stands for the angular distance between the source and the optical axis as seen from the lens. In real observations, this angle should be of the order $\gamma \sim D^{-1}_{ OL }$. In the pannel of Fig. \ref{fig:thetainfi} we depict these four quantities as a function of the parameter $l$. We see that the asymptotic position of the images $\vartheta_{\infty}$ and the delay time $\Delta T_{2,1}$ show a linear decrease in $l$. The distance between the first image and the other images $s$ also decreases, while $r_m$ increases, which means that the brightness of the first image becomes more intense in relation to the others.

\section{Conclusion \label{sec:conclu}}

In this article, we studied the gravitational lenses generated by a black hole described by the Kalb-Ramond (KR) solution, which is a Schwarzschild-type geometry with spontaneous Lorentz symmetry breaking implemented via a single new parameter $l$. This is the second paper in a series dedicated to constraining such a parameter using data from the Sagittarius $A^{\star}$ supermassive black hole at the heart of our Milky Way galaxy. In the first such article, we discussed orbital precession and managed to restrict $l$ to the interval:  $-0.18502 < l <0.06093 $ \cite{Ednaldo}, while in this one we used these constraints to study gravitational lensing in the same theory.

We first computed the exact expression of the deflection angle $\alpha$ in the framework of the KR geometry for massive particles at a finite distance of both observer and source using elliptic integrals. This result can be easily converted into the special cases of infinite distance when the inverse of the distance radius turn to zero, $u_S \rightarrow 0$ and $u_R  \rightarrow 0$, and for light when $v  \rightarrow 1$. We next used Taylor series expansions to obtain approximate expressions for the deflection angle in the weak field regime at both infinite and finite distances for massive particles with velocity $v$ and for massless particles (corresponding to $v=1$). The expressions found this way naturally generalize their Schwarzschild counterparts of the massless case (the latter found when $v=1$ and $l=0$). In the strong field regime of gravitational lensing we computed the deflection angle of light using the formalism developed by Bozza in \cite{Bozza:2002zj}, and found the coefficients of the logarithmic approximation \eqref{eq:explog} in Eqs. \eqref{eq:bmsf}, \eqref{eq:b1sf}, and \eqref{eq:b2sf}. We found that the  parameter $l$, within the constraint above, tends to reduce $\alpha$.

Based on the results above for the expressions of the deflection angle, we used the experimental data of the supermassive black hole at the center of our Galaxy, Sgr $A^{\star}$, shown in Table \ref{tab:tabela1}, and computed the observables associated with the weak and strong field limits of the gravitational lensing effect, using the formalism developed in \cite{Keeton:2006sa,Bozza:2002zj}. In the weak-field regime, we studied the behaviour of the angular separation $P_t$ \eqref{eq:P_t}, the difference of the angular positions $\Delta P$ \eqref{eq:deltap}, the total flux $F_t$ \eqref{eq:fluxt}, the difference of the fluxes $\Delta F$ \eqref{eq:deltaf}, the center of gravity $\Theta_{cent}$ \eqref{eq:thetacent} and the differential time delay $\Delta \tau$ \eqref{eq:deltatau}. We arrived at the conclusion that all these values can be measured with the current experimental instruments, but that the difference between the Kalb-Ramond solution and the Schwarzschild solution for $\beta <0.1$ is probably too small to be measured with current technology, unless the source is practically aligned with the lens. In the strong-field limit, we focused on the asymptotic position given by a series of images $\vartheta_{\infty}$ \eqref{eq:thetainfity}, the distance between the first image $\vartheta_1$ and all the others $s$ \eqref{eq:s}, the ratio between the flux of the first image and the flux of all other images $r_m$ \eqref{eq:r_m}, and in the time delay between a photon with 2 loops and a photon with one loop around the lens $\Delta T_{2,1}$ \eqref{eq:delaysf}.
In principle, the only observable that is within reach of actual measurements this time is the asymptotic position of the images, which varies between $23 \mu$ arcs and $28 \mu$ arcs.

The analysis carried out in this work only involves gravitational lensing associated to a distance source to the lens object, while we plan to further explore our setting in order to explore gravitational lensing of KR geometry when the main source of illumination is provided by the accretion disk. This way, we shall explore several aspects of shadow and photon ring images of these geometries and their comparison with those images cast by usual Schwarzschild black holes in looking for observational discriminators between them that can be searched for using very long base line interferometry.

\acknowledgments

MER  thanks Conselho Nacional de Desenvolvimento Cient\'ifico e Tecnol\'ogico - CNPq, Brazil, for partial financial support. This study was financed in part by the Coordena\c{c}\~{a}o de Aperfei\c{c}oamento de Pessoal de N\'{i}vel Superior - Brasil (CAPES) - Finance Code 001. FSNL acknowledges support from the Funda\c{c}\~{a}o para a Ci\^{e}ncia e a Tecnologia (FCT) Scientific Employment Stimulus contract with reference CEECINST/00032/2018, and funding through the research grants UIDB/04434/2020, UIDP/04434/2020, CERN/FIS-PAR/0037/2019 and PTDC/FIS-AST/0054/2021. DRG is supported by the Spanish Agencia Estatal de Investigación Grant No. PID2022-138607NB-I00, funded by MCIN/AEI/10.13039/501100011033, FEDER, UE, and ERDF A way of making Europe.



\end{document}